\documentclass{aa}

\usepackage{graphics}

\begin{document}


\title{Infrared emission towards \object{SN 1987A} 11 years after outburst:
Properties of the circumstellar dust
}
\author{Jg. Fischera \and R. J. Tuffs \and H. J. V\"olk}

\offprints{J\"org Fischera}
\institute{Max-Planck-Institut f\"ur Kernphysik, Saupfercheckweg 1, 69115 Heidelberg}

\mail{Joerg.Fischera@mpi-hd.mpg.de}

\date{Received  ---/ Accepted ---}

\authorrunning{Fischera et al.}
\titlerunning{IR emisison towards SN 1987A 11 years after outburst: 
Properties of the circumstellar dust}


\abstract{
Detailed models are presented for the late epoch mid infrared (MIR) emission 
from collisionally heated grains in the shocked circumstellar gas around SN 1987A.
Thermal dust emission from a region of moderate 
density interior to the thick inner ring seen with the Hubble Space Telescope (HST) 
is found to be a natural explanation for the MIR spectral energy distribution
measured by ISOCAM. The MIR-spectrum can be reproduced by a mixture of silicate-iron 
or silicate-graphite grains or by a composition of pure graphite grains. A composition
of pure iron grains on the other hand can be excluded and a pure silicate composition
does not seem to be very likely.
The dust-to-gas ratio in the interaction zone is $\sim 0.01\%$, an order of magnitude
lower than estimates for dust abundances in the winds of red supergiant (RSG) stars in the LMC.
This low dust abundance can be accounted for by a combination of 
evaporation through the UV-flash from the supernova outburst and subsequent
sputtering in the shocked gas.
For this explanation to hold, dust in the pre-supernova circumstellar medium (CSM)
would have to have been predominantly composed of grains other than graphite, with
a maximum size smaller than $\sim 0.1~\mu{\rm m}$. 
}

\maketitle


\section{Introduction}

SN 1987A, a supernova of Type II, made it possible for the
first time to make detailed observations of the interaction of a
supernova ejecta with the very innermost region of its CSM. 
IR measurements enable us to analyse the abundance, composition, and
size distribution of the circumstellar dust grains and to study grain 
destruction processes such as evaporation
by the UV-flash of the supernova outburst and sputtering in the shocked 
gas behind a very strong shock. From this, new insights 
into the history of the progenitor star can be obtained, providing a more
complete picture of the SN~1987A.

SN~1987A is the first supernova for which the progenitor was observed
prior to outburst. It has been identified as the most luminous star of the system 
Sanduleak $-69\degr 202$ in the Large Magellanic Cloud (LMC) (West, \cite{West87}) 
which had been classified as a blue supergiant (BSG) of spectral Type B3~I 
(Rousseau, \cite{Rousseau78}). 
Observations with the HST showed a complex axis symmetrical structure of its CSM with a thick
inner ring with a diameter of $\sim 1\farcs 6$ and two outer rings with
larger diameter at each side seen with a viewing angle of $\sim 43\degr$ (Burrows et al., \cite{Burrows95}). 
The analysis of dust scattered light of the supernova outburst suggest that the rings are connected
with gas and dust distributed in an hour glass - shaped shell (Crotts et al., \cite{Crotts95}). 

Initially, the supernova ejecta freely expanded in the thin wind of the
BSG, driving a blast wave into the CSM with a
velocity of $\sim 30\,000$~km/s.
The reappearance of the radio emission (Staveley-Smith et al., \cite{Staveley-Smith92}) almost coincidentally
with the appearance of soft X-ray emission detected by ROSAT (Beuermann et al., \cite{Beuermann94};
Gorenstein et al., \cite{Gorenstein94}) indicated
that after $\sim 1200$ days the blast wave had reached denser material
interior to the thick inner ring, which
slowed down the shock velocity to $2900\pm 480$~km/s (Gaensler et al., \cite{Gaensler97}).
Chevalier \& Dwarkadas (\cite{Chevalier95}) supposed that this denser gas is comprised of
material from a RSG phase of the progenitor of SN~1987A
before it evolved into the BSG that finally exploded. 
They refered to this region, 
in which the gas should be ionised through the photon flux of the progenitor star of SN 1987A, 
as the ``HII-region''.
The interaction between the fast thin wind of the BSG and the slow
moving thick wind of a RSG is also thought to 
be responsible for the larger structure of the CSM (Blondin \&
Lundqvist, \cite{Blondin93}; Martin \& Arnett, \cite{Martin95}).



In a previous paper (Fischera et al., \cite{Fischera02}, paper~I) we
presented  MIR measurements, made with ISOCAM (Cesarsky et al.,
\cite{Cesarsky96}) towards SN~1987A 11 years after the outburst. 
These reveal the central region around the supernova position as a resolved MIR source
with an extension and orientation consistent with the elliptical projection of the thick inner ring,
suggesting that the MIR emission is mainly circumstellar in origin. 
We found this emission is most probably from dust,
collisionally heated in the shocked gas downstream of
the blast wave as it expanded into the material of the HII-region interior to the thick inner ring.
On a purely energetical basis, all the emission could be from
condensates in the metal rich core region of the 
expanding ejecta (Fischera, \cite{Fischera00}). However, as argued in 
paper~I, an emission which  arises mainly from condensates does not
explain the measured extension or orientation of the MIR source.

In this paper we present detailed calculations of emission from grains in the shocked CSM
to analyse the implications of the ISOCAM measurements for the abundance, composition
and size distribution of the circumstellar grains.
In Sect. \ref{theorysection} we describe the dust model used to analyse the dust properties 
in the shocked CSM. Quantitative results 
are presented in Sect. 3. 
In Sect. 4 we discuss probably the most important destruction processes, which are evaporation
of grains through the UV-flash and sputtering in the shocked gas downstream of the blast wave.
In Sect. 5 this is used to derive information on grain abundance, and composition in the CSM prior to
the supernova outburst. A summary is given in Sect. 6.
As in paper~I we will 
assume a distance of 51 kpc to the supernova.

\section{Modelling the MIR emission from the shocked CSM}

\label{theorysection}
We consider grains that are collisionally heated in the shocked circumstellar plasma
downstream of the blast wave.

\begin{figure}[htbp]
  \resizebox{\hsize}{!}{\includegraphics{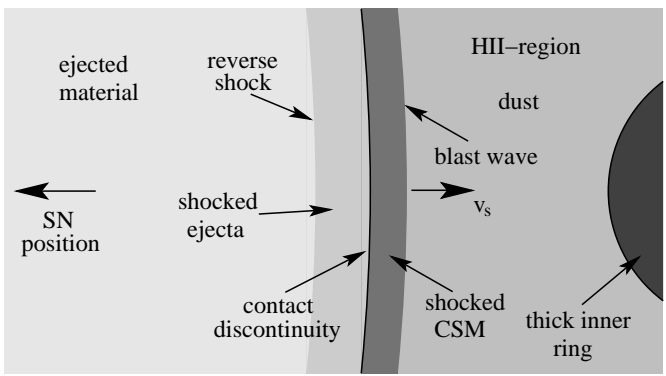}}
   \caption{
    \label{cartoon_shock} Cartoon of the basic structure caused by
	the interaction of the expanding ejecta with the HII-region.}
\end{figure}

The structure caused by the interaction of the expanding ejecta with
its CSM is visualised in figure \ref{cartoon_shock}.
The shocked circumstellar gas is compressed to a thin layer 
that grows approximately as $0.1v_{\rm S}t$ (Chevalier, \cite{Chevalier82}), where 
$v_{\rm S}$ is the assumed constant speed of the blast wave and $t$ the time since
the shock reached the HII-region. Downstream it is bounded by the contact discontinuity 
to the shocked outer parts of the expanding ejecta. This ejected material is heated 
by both the reverse shock and by a shock reflected from the inner boundary of the 
HII-region (Borkowski et al., \cite{Borkowski97}). It is thought
to be the origin of the strong emission in Ly$\alpha$ and H$\alpha$ at a distance of
$\sim 0.6$ of the radius of the thick inner ring (Michael et al., \cite{Michael98}), 
that was measured by the HST close to the time of 
the ISOCAM observations.
We do not expect strong MIR emission from
the shocked ejected material for the following reasons:
\begin{enumerate}
\item The material at the reverse shock is mainly hydrogen and 
has such a low abundance in metals 
(e.g. Woosley, \cite{Woosley88}), that, if any, only very few grains may have been formed
in this region of the ejecta. 
\item Further, as we show later (see Sect. \ref{graincompression}), the grain charge and the strength 
of the magnetic field in the 
shocked CSM in the interaction zone
makes it unlikely, that an observable amount of 
circumstellar grains have been overtaken by the outer parts of the ejecta. 
\item
The gas density of the ejecta at the reverse shock is, following
the numerical results from Borkowski et al. (\cite{Borkowski97}), much less
than the density downstream of the blast wave. In consequence, the heating of any ejected grains
at the reverse shock should be much weaker than for circumstellar grains downstram of the blast wave. 
\end{enumerate}



\begin{figure*}[tbh]
 \resizebox{0.49\hsize}{!}{\includegraphics{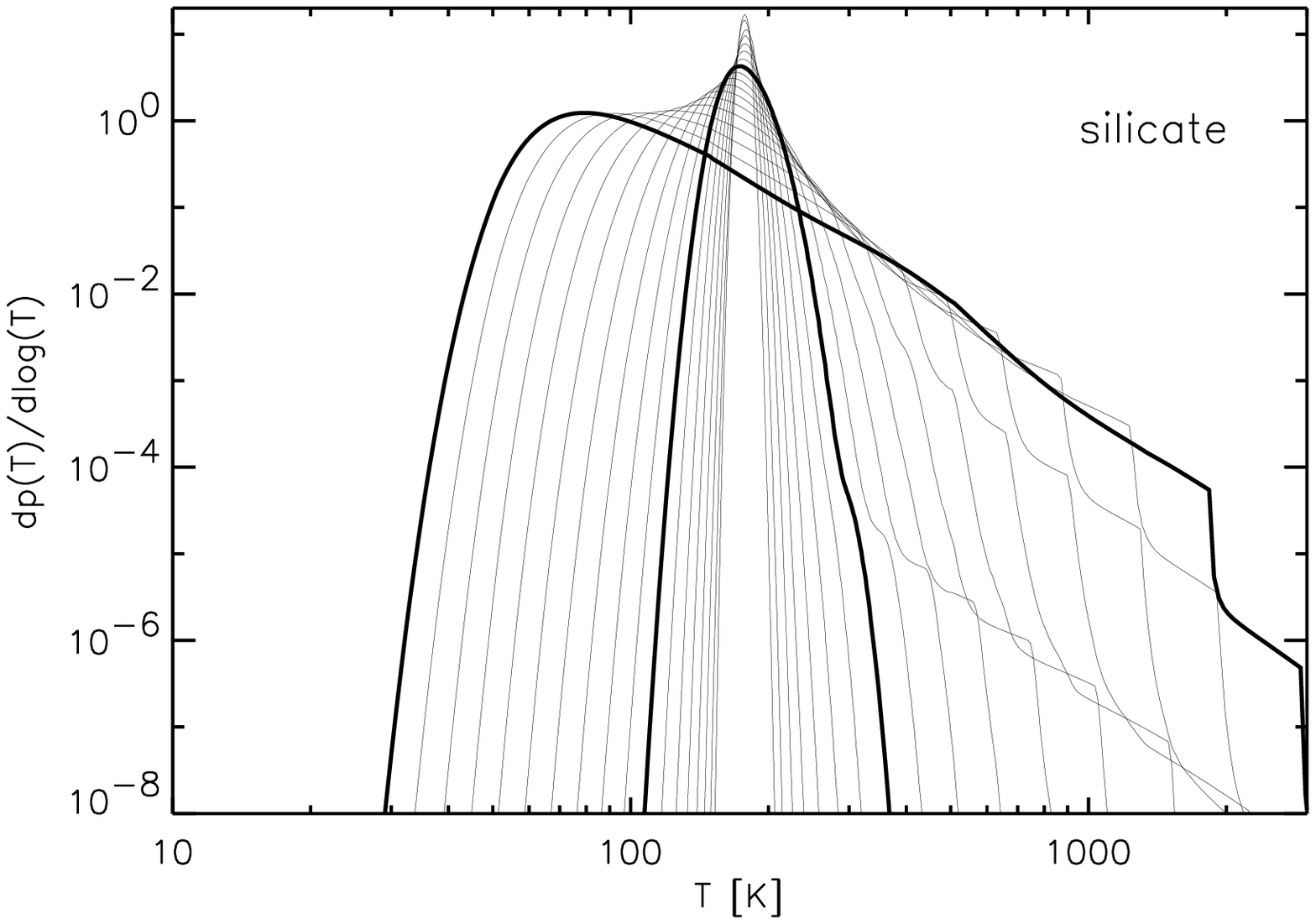}}
  \hfill
 \resizebox{0.49\hsize}{!}{\includegraphics{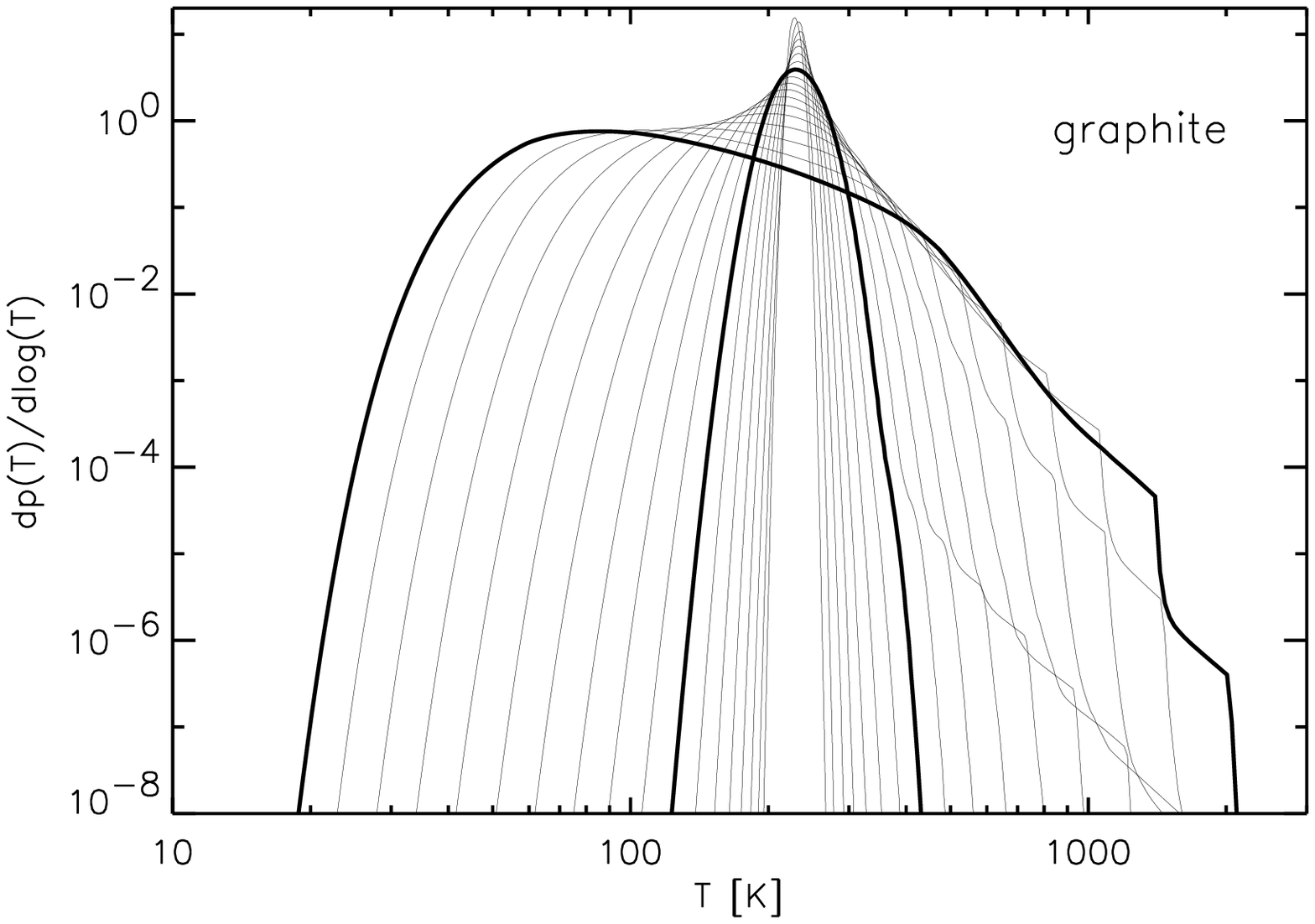}}
  \caption{
    \label{PT-curves}
    Temperature distributions of spherical silicate and graphite grains for various
    grain sizes in the shocked CSM, 
    assuming a density of the shocked gas of $n_{\rm H}=300~{\rm cm^{-3}}$ (model I).
    The variation in size between two lines is $d\log(a[\mu{\rm m}])=0.1$.
    The temperature distributions, shown with thicker lines, belong
    to grains with a radius of $0.001~\mu{\rm m}$ and $0.01~\mu{\rm m}$.
    The steps in the $p(T)$-curves are due to the approximation used
    for ion heating and appear roughly at temperatures where the thermal  
    energy of the grains is equal to the threshold energy of the ions 
    (see also Popescu et al., \cite{Popescu00}).}
\end{figure*}

\subsection{Gas parameters}

\label{parametergas}
The collisional heating of a grain in a hot plasma depends mainly on the temperatures
of electrons and ions and the number density of the different species of the gas.\footnote{
In general the heating, in particular the heating by the ions, has also a dependency 
on the relative motion of the grains in comparison to the motion of the gas, which is neglected for simplicity.} 
We fix these properties to be
broadly consistent with the observed radio and X-ray emission from the CSM. An analysis of
the X-ray observations made with the ROSAT satellite towards SN~1987A up to and including
the epoch of the ISOCAM observations is given in appendix~\ref{x-ray-flux}.
For temperatures above $\sim10^7\,{\rm K}$, as encountered
in the shocked gas, the grain heating of the smaller grains principally contributing
to the emission observed by ISOCAM is mainly determined by the plasma density (see e. g. Dwek \&
Arendt 1992). As the physical conditions are somewhat uncertain we will consider two cases 
(model I and II) with different temperatures and densities.
The parameters of model I and II are summarised in 
Table~\ref{gasparameter} and will be justified in the following.

\begin{table}[htbp]
  \caption{
    \label{gasparameter}
    Parameters of the shocked gas}
  \begin{tabular}{l||l|l}
  \hline
    & model I & model II \\
  \hline
  shock velocity $v_{\rm S}$      & 4100 km/s                 & 2900 km/s \\
  electron temperature $T_{\mathrm{e}}$ & $2\times 10^7\,\mathrm{K}$ & $2\times 10^7\,\mathrm{K}$ \\
  ion temperature $T_{\mathrm{i}}$ & $6\times 10^8\,\mathrm{K}$ & $3\times 10^8\,\mathrm{K}$ \\
  hydrogen density $n_{\mathrm{H}}$ & $300\,\mathrm{cm^{-3}}$ & $600\,\mathrm{cm^{-3}}$ \\
  helium density $n_{\mathrm{He}}$ & $2.5\,(n_{\mathrm{He}})_{\sun}$ &  $2.5\,(n_{\mathrm{He}})_{\sun}$ \\
  metallicity $Z$ & $0.3\,Z_{\sun}$ & $ 0.3\, Z_{\sun}$ \\
  \hline
  \end{tabular}
\end{table}

Borkowski et al. (\cite{Borkowski97}) have shown that an assumption of constant density, as suggested by
Chevalier \& Dwarkadas, can explain the soft X-ray emission (Hasinger et al., \cite{Hasinger96} ) until
at least 3000 days after outburst. They derived a pre-shock hydrogen density of $75~{\rm cm^{-3}}$ with a 
corresponding shock speed $v_{\rm S}$ of the blast wave of $4100~{\rm km/s}$. 
In their calculations the HII-region was modelled as a thick torus in which
the inner ring seen by the HST is embedded. They also mentioned that for a different structure of the HII-region 
higher densities would be possible. Indeed, a shock speed of $2900\pm 480~{\rm km/s}$ as derived from radio
observations (Gaensler et al., \cite{Gaensler97}) would require a hydrogen density of the HII-region of 
about $150~{\rm cm^{-3}}$. It might also be possible, that the density increased with time 
(see appendix~\ref{x-ray-flux}) although the X-ray
measurements are also consistent with a constant external density.

For simplicity we assume that at the time of the ISOCAM observations 
the blast wave was still expanding into a homogeneous HII-region. We will
consider hydrogen densities of $75~{\rm cm^{-3}}$ (model I) and $150 ~{\rm cm^{-3}}$ (model II) for the 
pre-shock region.
The density of the gas is assumed to be compressed by a factor of four downstream of the blast wave and 
to stay constant in the whole interaction zone. 

In converting the densities given by Borkowski et al. into hydrogen number density
and ionic abundances needed for the calculation of dust heating (Sect. \ref{modeldust}) we adopted 
the abundances of the thick inner ring given by
(Lundqvist \& Fransson, \cite{Lundqvist96}). On this basis we took
Helium to be a factor 
of 2.5 more abundant than in the sun, and approximated the metallicity 
of all heavier elements to be 0.3 solar.


Because of the long equipartition time of the shocked circumstellar plasma the electron
temperature should be much lower than the temperature for the ions and 
should increase with the distance to the outer shock (see e.g. Burrows et al, \cite{Burrows00}). 
Thus, in calculating the grain heating we
use two different temperatures for electrons and ions. As for the density, the
temperature of the shocked CSM is taken to be independent of position. 


The ion temperature $T_{\rm i}$ is taken to be (see e.g. Longair, \cite{Longair97})
\begin{equation}
  T_{\rm i}=2\,\frac{\gamma-1}{\left(\gamma +1\right)^2}\,\frac{\mu \,m_{\rm H}}{k_{\mathrm{B}}}\,v_{\rm S}^2
\end{equation}
with $\gamma=5/3$ and an atomic weight of $\mu=1.6$ appropriate to the abundances in the CSM.


For the electrons we choose a temperature close to those derived from X-ray observations and
predicted by numerical calculations appropriate to SN 1987A.
From X-ray observations made with the ROSAT-satellite Hasinger et al. (\cite{Hasinger96}) 
derived a temperature of approximately $T_{\rm e}\approx 1.2\times 10^7\,{\rm K}$. 
Analysing the X-ray spectrum of SN 1987A, taken later with CHANDRA, 
using a shock model with a constant $T_{\rm e}$, a higher electron temperature of the
order of $\sim 3.5\times 10^7\,{\rm K}$ was found
(Burrows et al. \cite{Burrows00}, Park et al., \cite{Park02}). In our calculation we will adopt
$T_{\rm e}=2\times 10^7$~K. 

\subsection{Calculation of the dust emission}

\label{modeldust}

The dust in the CSM is taken to be spherical for simplicity as there
is no observational evidence for other grain shapes in the CSM of
SN~1987A. Spherical grains are also generally  assumed in the literature
for grains produced in stellar winds (e.g. Gail \& Sedlmayr, \cite{Gail99}). 
However, if the grains are not too different from being spherical this assumption should not
influence significantly our results about the properties of the grains in the CSM.
For example, it has been found, that the temperature of spheroidal grains
heated by the interstellar radiation field (ISRF) can only vary by more than 10\%
if the axial ratio exceeds a value of 2 (Voshchinnikov et al., \cite{Voshchinnikov99}).

The grains are heated by collisions with electrons and ions of the hot ionized shocked plasma. 
The gas species is assumed to transfer all 
(if it sticks) or only a part (if it is not stopped) of the kinetic energy into thermal 
energy of the grain. The calculation of the energy deposition of
non-stopping  particles is based on their stopping-distances (ranges) in
solids\footnote{ The stopping power is due to ionisation losses in case
of electrons.  In case of the ions at energies considered
here the stopping power is partly due to interaction with the electrons and partly due to
inelastic scattering with the nucleons of the target material.}.
For the heating by ions we considered in addition to 
hydrogen and helium also the next most abundant elements oxygen, nitrogen and carbon.
The emission from larger grains we derived from their equilibrium temperatures. 
For smaller grains, where the deposited
energy is typically larger than the thermal energy, we took their temperature fluctuations into account.
The model for stochastic dust emission from a hot plasma
is described in more detail in Popescu et al. (\cite{Popescu00}).
As typical dust species we consider graphite and silicate grains.
The temperature distributions of very small silicate and graphite grains in model I
are shown in Fig.~\ref{PT-curves} where it can be seen, that small grains 
can be heated to very high temperatures. The smallest grains with $10$~\AA{} radius 
will reach temperatures well above their evaporation temperature. 
The grains contributing to the measured IR emission should therefore not be smaller than this size.

Because iron is expected to form in cool stellar outflows independent of the C/O ratio and 
is potentially one of the main condensates in the circumstellar environment of oxygen rich stars 
(see e.g. Gail \& Sedlmayr, \cite{Gail99}, and references there),
we also carried out calculations for pure iron grains.
For electrons in iron grains 
we used the analytical expression for the electron range in graphite derived by Dwek 
\& Smith (\cite{DwekSmith96}) on basis of observational data, correcting for the different
density. The optical properties
of iron spheres we derived using MIE-theory (Bohren \& Huffman, \cite{Bohren83}), whereby we
included the dependence of the dielectric function
in the IR on size and temperature of the iron grains (Fischera, \cite{Fischera00}). 
For the heat capacity of iron grains cooler than 298~K 
we used values tabulated in the American Institute of Physics Handbook (\cite{AIP72}).
For iron grains warmer than this we used an analytical expression for the heat capacity given
by Chase (\cite{Chase98}), assuming that iron is in the $\alpha$-$\delta$-phase.

\subsubsection{Grain temperatures in the shocked CSM}

\begin{figure}[htbp]
 \resizebox{\hsize}{!}{\includegraphics{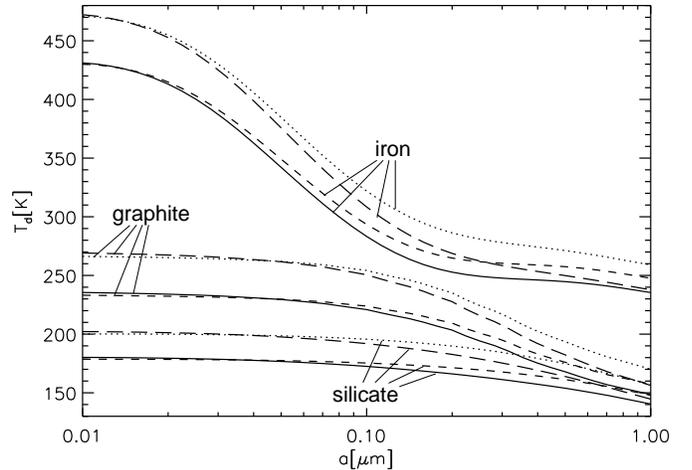}}
   \caption{
    \label{equil_temp_hii}
    Equilibrium temperatures of spherical silicate, graphite and iron grains as a function of grain radius $a$
    in the shocked gas downstream of the blast wave. The grain temperatures in model I and II are represented 
    by the solid and the long dashed line, respectively. Also shown is the effect of raising 
    $T_{\rm e}$ from $2\times 10^7$~K to $3.5\times 10^7$~K 
    (short dashed and dotted lines, respectively, for models I and II).}
\end{figure}

The variation of the equilibrium temperature of spherical
silicate,  graphite, and iron grains in the shocked CSM is shown
in Fig.~\ref{equil_temp_hii}. The different behaviour of the three species
allows some information about dust composition to be derived from the ISOCAM
measurements.
Grain temperatures for the case $T_{\rm e}=3.5\times 10^7\,{\rm K}$ 
are also given.
As seen in the figure the grain
temperature of small silicate and graphite grains
is nearly insensitive to $T_{\rm e}$
since the electrons are not stopped by the grains. The equilibrium temperatures
are consistent with the colour temperatures $T_{\rm c}$ we derived in 
paper~I for a modified Planck spectrum $F_{\lambda}\propto
\lambda^{-\beta}B_{\lambda}(T_{\rm c})$, where $B_{\lambda}(T_{\rm c})$
is the Planck function, ranging from $\sim 200$~K  ($\beta=2$) to $\sim
290$~K ($\beta=0$). Because of stronger heating due to the higher grain
density and the weaker cooling due to lower emissivities in comparison to
silicate and graphite grains pure iron grains attain the highest
temperatures. We also note that for the  conditions in the shocked CSM,
graphite grains are hotter than silicate grains. In less dense plasmas (e.g. $n_{\rm H}=1~{\rm cm^{-3}}$) 
one would expect a slightly
higher temperature for silicate grains (Dwek, \cite{Dwek87}, Fischera, \cite{Fischera00}).

The temperature of small spherical iron grains depends strongly on grain size. 
This was also noted by Chlewicki \& Laureijs (\cite{Chlewicki88}) for iron grains
heated by the interstellar radiation field.



\subsubsection{Grain size distribution in the CSM}

In order to construct a SED for comparison with the data a functional form
for the grain size distribution must be adopted.
The size distribution of the grains in the shocked circumstellar environment of SN 1987A 
is not known and might be modified in comparison to the initial distribution in the CSM
by different processes like evaporation during the UV-flash or sputtering in the
shocked gas as will be shown later. 
However, for simplicity we assume, that the grains
in the shocked CSM have a grain size distribution similar to grains
in the ISM of our galaxy and can be described by
a simple power law $\mathrm{d}n\propto a^{-k}\,\mathrm{d}a$ with power index $k$
and a minimum and a maximum grain size $a_{\rm min}$ and $a_{\rm max}$.
Following Biermann \& Harwit (\cite{Biermann80})
a power law distribution should be
a general description of a grain size distribution resulting 
from grain-grain collisions and would especially describe
the grain size distribution in the atmospheres of red-giants.
 In contrast to the emission from grains in the ISM, heated by the ISRF, 
where only stochastically heated small grains
can achieve high enough temperatures to emit at shorter wavelengths 
(Draine \& Anderson, \cite{Draine85}),
in the shocked CSM of SN 1987A larger grains can also contribute to the measured fluxes.
The composite SEDs are relatively insensitive to $a_{\rm min}$, which we fixed at
10~\AA{}.

\subsubsection{Free dust model parameters}


For each of model I and model II for the shocked gas we made
calculations taking various combinations of the following 
parameters as free variables:
\begin{itemize}
  \item $M_{\rm d}$, the dust mass. This was a scaling parameter in all calculations.
  \item $k$, the exponent in the grain size distribution ${\rm d}n(a)\propto a^{-k}\,{\rm d}a$.
  \item $a_{\rm max}$, the maximum grain size.
  \item Grain composition. Five compositions were considered: Pure silicate, pure graphite, pure iron,
    a silicate-iron mixture and a silicate-graphite mixture. The relative composition in the mixtures
    was considered as a free variable.
\end{itemize}
The most probable values for the variables were found through a 
$\chi^2$-fit to the measured flux densities, taken from paper~I,
which we colour corrected (see Blommaert et al., \cite{Blommaert01}) 
on the basis of the modelled spectrum. 
In addition we derived for each fit the luminosity $L_{\rm d}$ of the theoretical dust emission spectrum.
The one sigma uncertainties of the parameters were calculated from the $\chi^2$-fit by varying 
$\Delta \chi^2_1=\chi^2_1 - \chi^2_{\mathrm{min}}$ until $\Delta\chi^2_1=1$ (see e.g. Press et al., \cite{Press92}). 
To estimate the uncertainties of the fitted parameters 
$a_{\mathrm{min}}$, $k$ and the dust mixture the dust mass $M_\mathrm{d}$ was taken to be a free variable.
For simplicity the uncertainty of $M_\mathrm{d}$ itself was derived with all other parameters fixed.

\section{Results}

\label{resultsection}
The results of the model calculations are summarised in Table~\ref{results-dustmodel}, where we give
for each fit also the reduced $\chi^2_{\nu}$, where $\nu$ is the number of free parameters.
The best fits obtained for the two mixtures are shown in Fig.~\ref{spectral-fit} together
with the fitted spectrum for silicate grains.
The emission spectrum is such that 
measurements at longer wavelengths (e.g. Lundqvist et al., \cite{Lundqvist99}) 
are too insensitive to probe the emission from this region.
The fits can be summarised as follows:

\begin{figure*}[htbp]
 \resizebox{0.49\hsize}{!}{\includegraphics{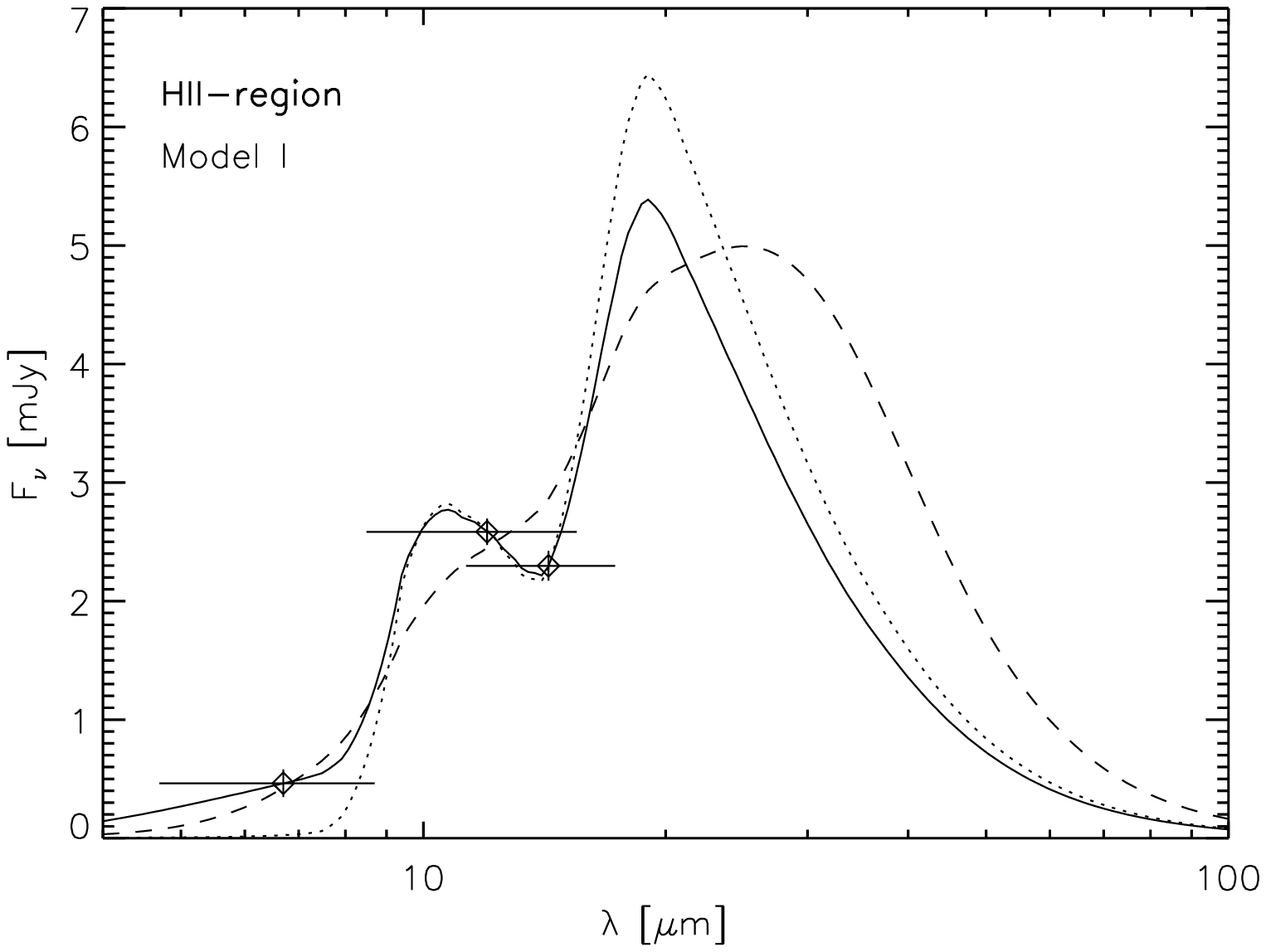}}
  \hfill
 \resizebox{0.49\hsize}{!}{\includegraphics{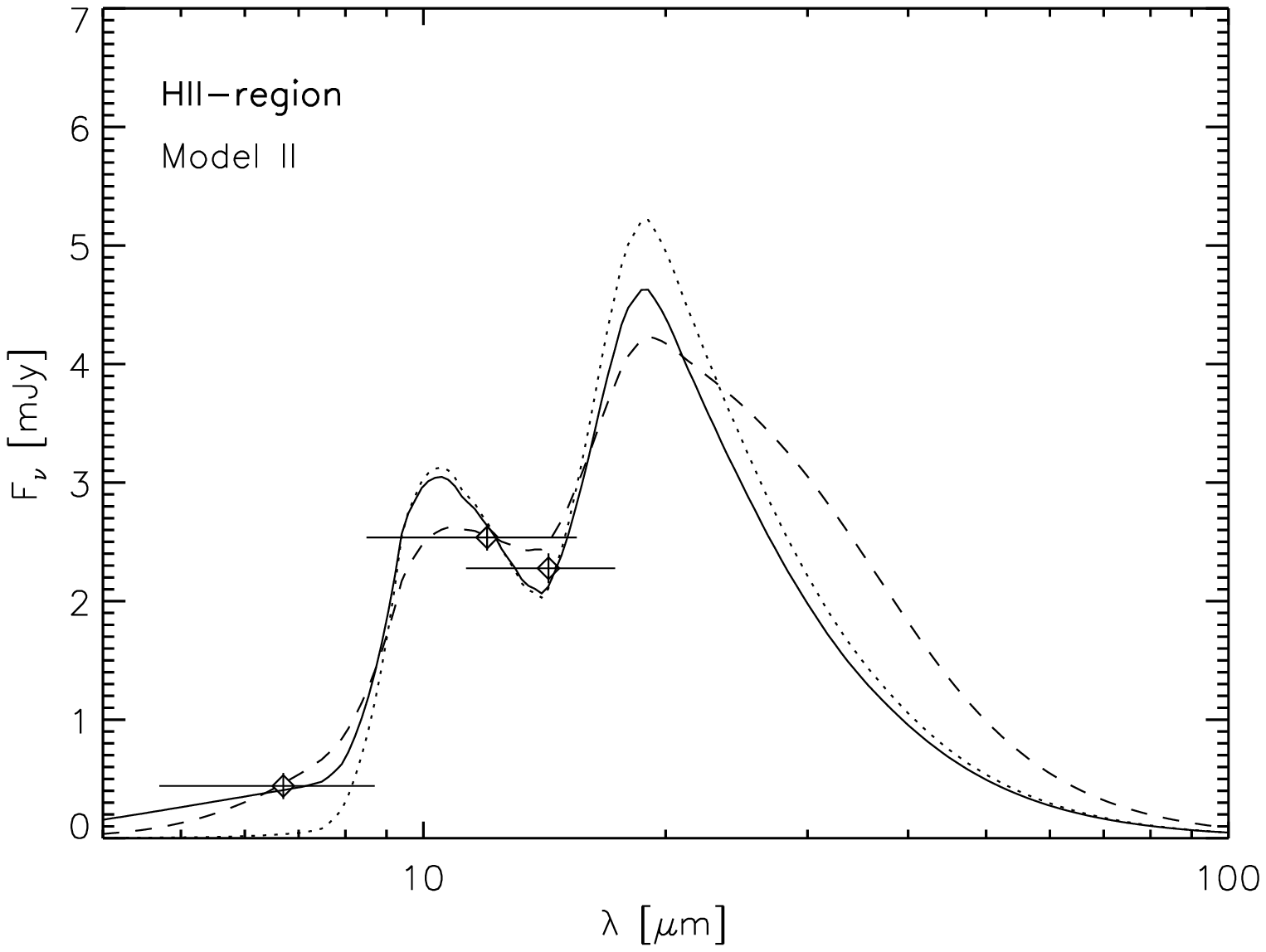}}
  \caption{
    \label{spectral-fit}
    Theoretical emission spectra of dust in the shocked circumstellar medium of SN 1987A for model I 
        ($n_{\rm H}=300\,{\rm cm^{-3}}$) and model II ($n_{\rm H}=600\,{\rm cm^{-3}}$) .
    Shown are spectra from pure silicate (dotted line) and the best fits with a 
        mixture of silicate with graphite (dashed line) and with a mixture of silicate with 
        iron (solid line). The grains are assumed to have a grain size distribution ${\rm d}n\propto a^{-3.5}\,{\rm d}a$
        with minimum and maximum grain radii of $10~$\AA{} and $0.25~\mu{\rm m}$.
        The ISOCAM fluxes (triangles), shown with the widths of the
        filters as horizontal lines, have been colour corrected
        to the spectrum of the silicate-iron mixture. The shown uncertainties are the
	absolute uncertainties in the integrated flux densities (taken from Table~2 of
 paper~I)
	which are $25\%$ ($6.75~\mu{\rm m}$), $4.3\%$ ($12~\mu{\rm m}$) and
	5.5\% ($14.3~\mu{\rm m}$). These uncertainties are a combination of the systematic 
	calibration uncertainties, which dominates at 12 and 14.3~$\mu{\rm m}$, and the
	random uncertainties.
    }
\end{figure*}

{
\setlength{\tabcolsep}{0.6mm}
\renewcommand{\footnoterule}{\rule{0mm}{0.mm}}
\begin{table*}[htbp]
 \begin{minipage}[t]{\hsize}
  \caption{
    \label{results-dustmodel}
    Results from the dust modelling.}
  \begin{tabular}{c c | c c c c  c  c | c c c c  c c }
    \hline
     & & \multicolumn{6}{c|}{results model I ($n_{\rm H}=300\,{\rm cm^{-3}}$)} & 
     \multicolumn{6}{c}{results model II ($n_{\rm H}=600\,{\rm cm^{-3}}$)} \\ 
    \hline
    \hline
    \multicolumn{2}{c|}{compos.} & mixture\footnote{Relative mass of silicate 
      to iron or silicate to graphite, respectively.} &
    $k$ & $a_{\rm max}[\mu{\rm m}]$ & $M_{\rm d} [10^{-6}\mathrm{M_{\sun}}]$ & $L_{\rm d}$[$10^{28}$W] & $\chi^2_{\nu}$ & 
    mixture & $k$ & $a_{\rm max}[\mu{\rm m}]$ & $M_{\rm d} [10^{-6}\mathrm{M_{\sun}}]$ & $L_{\rm d}$[$10^{28}$W] & 
    $\chi^2_{\nu}$\\
    \hline
    \multicolumn{14}{c}{single variable: $M_\mathrm{d}$ ($\nu=2$)}\\
    \hline
       \multicolumn{2}{l|}{silicate} & --- & 3.5 & 0.25 & $0.97\pm 0.03$\footnote{The uncertainty of
         $M_{\rm d}$ was derived with all other parameters fixed.}  & $2.94$  & 6.1 & 
         --- & 3.5 & 0.25 & $0.50\pm 0.02$ & $2.63$ & 5.3 \\
       \multicolumn{2}{l|}{graphite} & --- & 3.5 & 0.25 & $1.04\pm 0.04$ & $3.35$  & 1.6 & 
         --- & 3.5 & 0.25 & $0.56\pm 0.02$ & $3.15$ & 3.5 \\
       \multicolumn{2}{l|}{iron}     & --- & 3.5 & 0.25 & $1.31\pm 0.05$ & $3.35$  & 55. & 
         --- & 3.5 & 0.25 & $0.86\pm 0.03$ & 3.65 & 81.\\
    \hline
    \multicolumn{14}{c}{two variables: $M_\mathrm{d}$ and $k$ ($\nu=1$)}\\
    \hline
       \multicolumn{2}{l|}{silicate} & --- & $3.99_{-0.24}^{+0.23}$ & 0.25 & $0.85\pm 0.03$ & 2.83 & 8.9 &
         --- & $3.39_{-1.32}^{+0.44}$ & 0.25 & $0.51\pm 0.02$ & 2.64 & 11. \\
       \multicolumn{2}{l|}{graphite} & --- & $3.56_{-0.61}^{+0.30}$ & 0.25 & $1.03\pm 0.04$ & 3.36 & 3.1 &
         --- & $<2.25$\footnote{The limit given for $k$ corresponds to $\chi^2(k)-\chi^2(k=0)=1$; here
        $M_{\rm d}$, $L_{\rm d}$ and $\chi^2_{\nu}$ were calculated for $k=0$.} & 0.25 
         & $0.67\pm0.02$ & 3.03 & 1.3\\
       \multicolumn{2}{l|}{iron}     & --- & $<0.53$ & 0.25 & $1.51\pm 0.05$ & 2.54 & 7.8 &
         --- & $<0.28$ & 0.25 & $1.09\pm 0.04$ & 2.59 & 18.\\
    \hline
    \multicolumn{14}{c}{two variables: $M_\mathrm{d}$ and dust mixture ($\nu=1$)}\\
    \hline
       \multicolumn{2}{c|}{si.+iron} & $2.09_{-0.61}^{+1.09}$ & 3.5 & 0.25 & $1.11\pm0.04$ & 3.18 & 0.0 &
         $2.33_{-0.73}^{+1.56}$ & 3.5 & 0.25 & $0.60\pm 0.02$ & 2.96 & 2.4 \\
       \multicolumn{2}{c|}{si.+iron} & $2.90_{-0.82}^{+1.57}$ & 3.5 & 0.10 & $0.97\pm0.03$ & 3.18 & 0.3 &
         $3.09_{-0.95}^{+2.09}$ & 3.5 & 0.10 & $0.51\pm 0.02$ & 2.95 & 4.4 \\
       \multicolumn{2}{c|}{si.+iron} & $3.35_{-0.95}^{+1.89}$ & 3.5 & 0.06 & $0.93\pm0.03$ & 3.19 & 0.75 &
         $3.39_{-1.05}^{+2.41}$ & 3.5 & 0.06 & $0.48\pm 0.02$ & 2.96 & 5.7 \\
       \multicolumn{2}{c|}{si.+gra.} & $0.25_{-0.25}^{+0.56}$ & 3.5 & 0.25 & $1.03\pm 0.03$ & 3.27 & 2.4 &
         $0.73_{-0.38}^{+0.73}$ & 3.5 & 0.25 & $0.53\pm 0.02$ & 2.94 & 0.2  \\
       \multicolumn{2}{c|}{si.+gra.} & $0.43_{-0.33}^{+0.64}$ & 3.5 & 0.10 & $0.93\pm 0.03$ & 3.22 & 1.07 &
         $0.84_{-0.40}^{+0.74}$ & 3.5 & 0.10 & $0.47\pm 0.02$ & 2.92 & 0.47  \\
       \multicolumn{2}{c|}{si.+gra.} & $0.52_{-0.34}^{+0.68}$ & 3.5 & 0.06 & $0.89\pm 0.03$ & 3.20 & 0.55 &
         $0.88_{-0.40}^{+0.75}$ & 3.5 & 0.06 & $0.45\pm 0.02$ & 2.91 & 1.45  \\
    \hline
    \multicolumn{14}{c}{two variables: $M_\mathrm{d}$ and $a_{\rm max}$ ($\nu=1$)}\\
    \hline
       \multicolumn{2}{l|}{silicate} & --- & 3.5 & $0.03_{-0.02}^{+0.07}$ & $0.78\pm 0.03$ & 2.78 & 9.3 &
         --- & 3.5 & $0.51_{-0.41}^{+1.93}$ & $0.60\pm 0.02$ & 2.70 & 10.2 \\
       \multicolumn{2}{l|}{graphite} & --- & 3.5 & $0.19_{-0.15}^{+0.30}$ & $1.01\pm 0.03$ & 3.36 & 3.1 &
         --- & 3.5 & $0.97_{-0.44}^{+1.60}$ & $0.75\pm0.03$ & 3.13 & 3.3\\
       \multicolumn{2}{c|}{si.+iron} & $2.09$ & 3.5 & $0.26_{-0.15}^{+0.53}$ & $1.12\pm0.04$ & 3.18 & 0.0 &
         $2.33$ & 3.5 & $1.55_{-1.24}^{>10.0}$ & $1.08\pm 0.04$ & 3.05 & 0.9 \\
       \multicolumn{2}{c|}{si.+gra.} & $0.25$ & 3.5 & $0.05_{-0.03}^{+0.14}$ & $0.90\pm 0.03$ & 3.28 & 1.0 &
         $0.73$ & 3.5 & $0.42_{-0.33}^{+0.71}$ & $0.59\pm 0.02$ & 2.95 & 0.05 \\ 
  \hline
  \end{tabular}
 \end{minipage}
\end{table*}
}

\begin{enumerate}
  \item Fit parameter: $M_{\rm d}$.\\
We simply compared the measured flux densities in the IR with the theoretical emission spectra
        of three different grain
    compositions (pure silicate, graphite or iron) with fixed grain size distribution for
    $a_{\rm max}=0.25~\mu{\rm m}$ and $k=3.5$ as proposed for the grains in the ISM (MRN, \cite{MRN77}).\\ 
    For both model I and model II a better fit is achieved under the assumption
    of pure graphite grains instead of silicate grains in the circumstellar environment.
    As seen in Fig.~\ref{spectral-fit} the flux at $6.7\,\mu{\rm m}$ 
    is too high to be emitted from silicate grains. 
    Because of the high predicted temperatures a pure iron composition is even more unlikely. 
  \item Fit parameter: $k$, $M_{\rm d}$.\\
For each of pure silicate, graphite and iron we varied the power $k$ of the size distribution with
        fixed $a_{\rm max}=0.25~\mu{\rm m}$.\\
    The derived grain size distributions for graphite grains in model I 
    and for silicate grains in both models are consistent with $k=3.5$.
    For graphite grains in model~II, which reach too high temperatures, a flatter distribution ($k<2.25$)
    is required. 
    No grain size distribution could be found which admitted a pure iron solution.
  \item Fit parameter: Dust mixture, $M_{\rm d}$.\\
    For a fixed grain size distribution with power $k=3.5$ and $a_{\rm max}=0.25~\mu{\rm m}$
    we varied the composition
    of circumstellar grains, for a mixture of silicate with iron grains and a mixture 
    of silicate with graphite grains.\\
    Excellent fits could be achieved for silicate-iron mixture in model~I and for the silicate-graphite 
    mixture in model~II (see also Fig.~\ref{spectral-fit}).
  \item Fit parameter: $a_{\rm max}$, $M_{\rm d}$.\\
    For a given dust composition (silicate, graphite and the two derived mixtures)
        and a fixed size distribution with power $k=3.5$ we varied $a_{\rm max}$.\\
    The maximum grain sizes in model I are significantly smaller than
    in model II. In model~I for 
    silicate, graphite or the silicate-graphite mixture a better fit can be
    achieved with 
    $a_{\rm max}<0.25~\mu{\rm m}$. 
    The estimated maximum grain sizes of all 
    considered compositions 
    in model II are several times larger. 
    We also examined a pure iron composition
    and found that the maximum grain size would be unrealistically large
    with $a>10~\mu{\rm m}$. 
\end{enumerate}

Due to the large number of free parameters several solutions
with different compositions and grain size distributions are possible. 
The derived values like $a_{\rm max}$, $k$ or the mixture of the composition 
are interrelated and depend on the plasma density.
Despite these uncertainties the following conclusions could be
reached largely independent of the assumptions:

\begin{enumerate}
\item{Luminosity}\\
For the luminosity of the collisionally heated dust $11$ years after outburst we obtained
(dependent on composition and the assumed plasma density)  
values for  $L_{\rm d}$ in the range  
$2.91\times 10^{28}$ to $3.35\times 10^{28}\,{\rm W}$. 
This can be compared with the crude estimate
of $L_{\rm d}=2.5\times 10^{28}$~W given in paper~I where we
approximated the dust emission spectrum with a simple modified black body spectrum.
As already mentioned in paper~I the luminosity in the MIR 
is larger than in X-rays. Despite the increase of the X-ray luminosity since detection $\sim 1400$
days after outburst (Hasinger et al. \cite{Hasinger96}) the luminosity in the energy range
0.5 to 10 keV even $\sim 13$ years 
after outburst was only $L_X\approx 2\times 10^{28}\,{\rm W}$ (Burrows et al., \cite{Burrows00}). 

\item{Composition}\\
The dust is most likely a mixture of silicate with
iron or silicate with graphite. Also pure graphite gives a reasonable fit. 
Pure silicate on the other hand seems to be unlikely and pure iron can be excluded.

\item{Dust-to-gas ratio}\\
Considering only cases where $a_{\rm max}$ is not larger than 
$0.25~\mu{\rm m}$ the dust mass is found to be nearly independent of
composition and  grain size distribution. 
The corresponding acceptable fits for model~I ($n_{\rm H}=300\,{\rm cm^{-3}}$) yield dust masses
in the range $(0.9~\mathrm{to} ~1.1)\times 10^{-6}~\mathrm{M_{\sun}}$ and those for 
model~II ($n_{\rm H}=600\,{\rm cm^{-3}}$) 
yield dust masses in the range $(0.45~\mathrm{to}~0.67)\times 10^{-6}~
\mathrm{M_{\sun}}$. 

Taking these values the dust-to-gas ratio is only mildly dependent
on gas density as one would also expect.
The mass of the X-ray emitting gas is given approximately by
\begin{equation}
  \label{gasmass}
  \mathrm{M_{gas}}\sim \frac{m_{\rm H}EM}{n_{\rm H}},
\end{equation}
where $EM$ is the emission measure from X-ray observations and $m_{\rm H}$ the hydrogen mass.
We have taken the emission measure from Hasinger et al.
(\cite{Hasinger96}) with $(1.4\pm 0.4)\times 10^{57}\,{\rm cm^{-3}}$ $\sim 2500$ days after outburst
and extrapolated this result with $t^{2.06}$ (see appendix \ref{x-ray-flux}) to 4000 days. 
Using equation \ref{gasmass} we derive a gas mass of 
$1.0\times 10^{-2}/n_{\mathrm H}[300~{\rm cm^{-3}}]~\mathrm{M_{\sun}}$.
Overall, 
this yields a dust-to-gas ratio in the range $(0.9~\mathrm{to}~ 1.3)\times 10^{-4}$.

If the grain size distribution is allowed to extend to grains with
radii much larger  than $0.25~\mu{\rm m}$
the acceptable fits imply that the dust-to-gas ratio could be as high as $2.2\times 10^{-4}$. 
As will be seen in Sect.~\ref{destr_section} the pre-supernova dust abundance would however
be almost unaffected.
\end{enumerate}





\section{Destruction processes due to the supernova}

\label{destr_section}

To compare the results for dust abundance and dust composition in the CSM of SN~1987A with
expectations for the CSM prior to the supernova event, account must be taken of the grain destruction
processes due to the supernova and its remnant.

\label{graincompression}
Before quantitatively discussing grain evaporation (section \ref{evaporationsection}) and grain sputtering 
(section \ref{sputteringsection}) we briefly 
demonstrate that the grains are dynamically coupled to the gas due to the betatron effect.
For example the potential of a graphite grain with $a=0.1~\mu{\rm m}$ at a temperature of $T_{\rm i}=3\times 10^8~{\rm K}$
is more than 30 Volt (Draine \& Salpeter, \cite{Draine79a}). 
The larmor radius of the grain is given by 
$R_{\rm L}=m_{\rm gr}v_{\rm gr}/Q_{\rm gr}B$, where $m_{\rm gr}$ is
the mass of the grain, $v_{\rm gr}$ the relative velocity of the grain to the gas (initially 
$\frac{3}{4}v_{\rm S}$), $Q_{\rm gr}$ the grain charge and $B$ the magnetic field.
For $B$ we adopt the value for equipartion between magnetic field
and relativistic particles (Longair, \cite{Longair97}), calculated from an extrapolation 
of the synchrotron emission after 1200 days (Gaensler et al., \cite{Gaensler97}). 
Assuming the radio emission to arise
from the same volume as the X-ray emission (see appendix~\ref{x-ray-flux}),
this yields $B\sim 10^{-7}$~T,
in agreement with the value found by Ball \& Kirk (\cite{Ball92}).
Taking the shock velocity to be $2900\,{\rm km/s}$, as derived from radio observations
(Gaensler et al., \cite{Gaensler97}), the larmor radius $R_{\rm L}$
is then of the order of $2.1\times 10^{11}~{\rm m}$, which is less then $\sim 10^{-4}$
of the diameter of the radio emission region.
By comparison, the distance of the outer blast wave to the 
contact discontinuity increases with $\sim 0.1\,v_{\rm S}\,t$
(Chevalier, \cite{Chevalier82}). Thus, the grains comoved with the 
gas after roughly one month.


To discuss the mass loss due to evaporation and sputtering we assumed, as in our earlier 
examination of the IR emission,
that the grains in the CSM at the time of the supernova explosion 
had a grain size distribution with $k=3.5$.
Again we choose as minimum grain size $a_{\rm min}=10\,$\AA{}.

\subsection{Evaporation during the UV-flash}

\label{evaporationsection}
The evaporation of grains in the neighbourhood of a supernova in general
has already been the subject of earlier examinations
(e.g. Draine \& Salpeter, \cite{Draine79b}, Draine, \cite{Draine81},
Pearce \& Mayes, \cite{Pearce86}) and also has been discussed for silicate grains for the SN 1987A 
(Emmering \& Chevalier, \cite{Emmering89}, Timmermann \& Larson, \cite{Timmermann93}). 
All considered the effect of evaporation
for single grain sizes only. Here
we are interested in the mass loss of grains with a certain 
size distribution in the shocked plasma behind the outer shock front. 
We 
use more recent theoretical results for the spectrum, duration and luminosity of the UV flash 
(Ensman \& Burrows, \cite{Ensman92}).
Apart from silicate we also consider iron and graphite grains.

As a reasonable
assumption, we only consider the evaporation of grains during this UV-flash,
when the grains reached their highest temperatures. For the luminosity, temperature, and
duration of the UV-flash we take
the theoretical results of the model \emph{500full1} from Ensman \& Burrows (\cite{Ensman92}).
For our calculation,
the spectrum is taken to be a simple black body spectrum with the given colour temperature from 
Ensman \& Burrows normalised to the luminosity.

To simplify the 
derivation of the mass loss of grains during the UV-flash,
we neglect the effect of stochastic heating
and assume that all grains are at their equilibrium temperatures. 
The evaporation of atoms from the surface of a grain with temperature $T_{\rm d}$ leads to a 
reduction in radius, that can be described by
\begin{equation}
  \label{evap}
  \frac{{\mathrm d}a[\mu{\mathrm{m}}]}{{\mathrm d}t[s]}\sim \frac{1}{3}\,10^{11}
  e^{-W(a)/k_{\mathrm{B}}T_{\rm d}(a)},
\end{equation}
where $k_{\rm B}$ is the Boltzmann constant and $W(a)$ the energy single atoms need to escape
from the surface, which depends on the radius of the grain because of the surface tension of the grain. 
The coefficient is chosen to be close to the value given by 
Guhathakurta \& Draine (\cite{Guhathakurta89}) and Voit (\cite{Voit91}) for graphite and silicate.
The energies $W(a)$ for silicate and graphite are adopted from Guhathakurta \& Draine. For iron we adopt
\begin{equation}
W(a)  =  (50\,000-22\,000\,N(a)^{-1/3})k_{\mathrm{B}},
\end{equation}
where $N(a)$ is the number of atoms in the grain.
Here we have chosen a bounding energy of the atoms close to $U=4.29~{\rm eV}$, 
given in Gerthsen et al. (\cite{Gerthsen89}),
and a surface tension of $\sigma=1.8~{\rm J/m^2}$, given in Lev\`efre (\cite{Lefevre79}). 

\begin{figure*}[htbp]
  \resizebox{0.49\hsize}{!}{\includegraphics{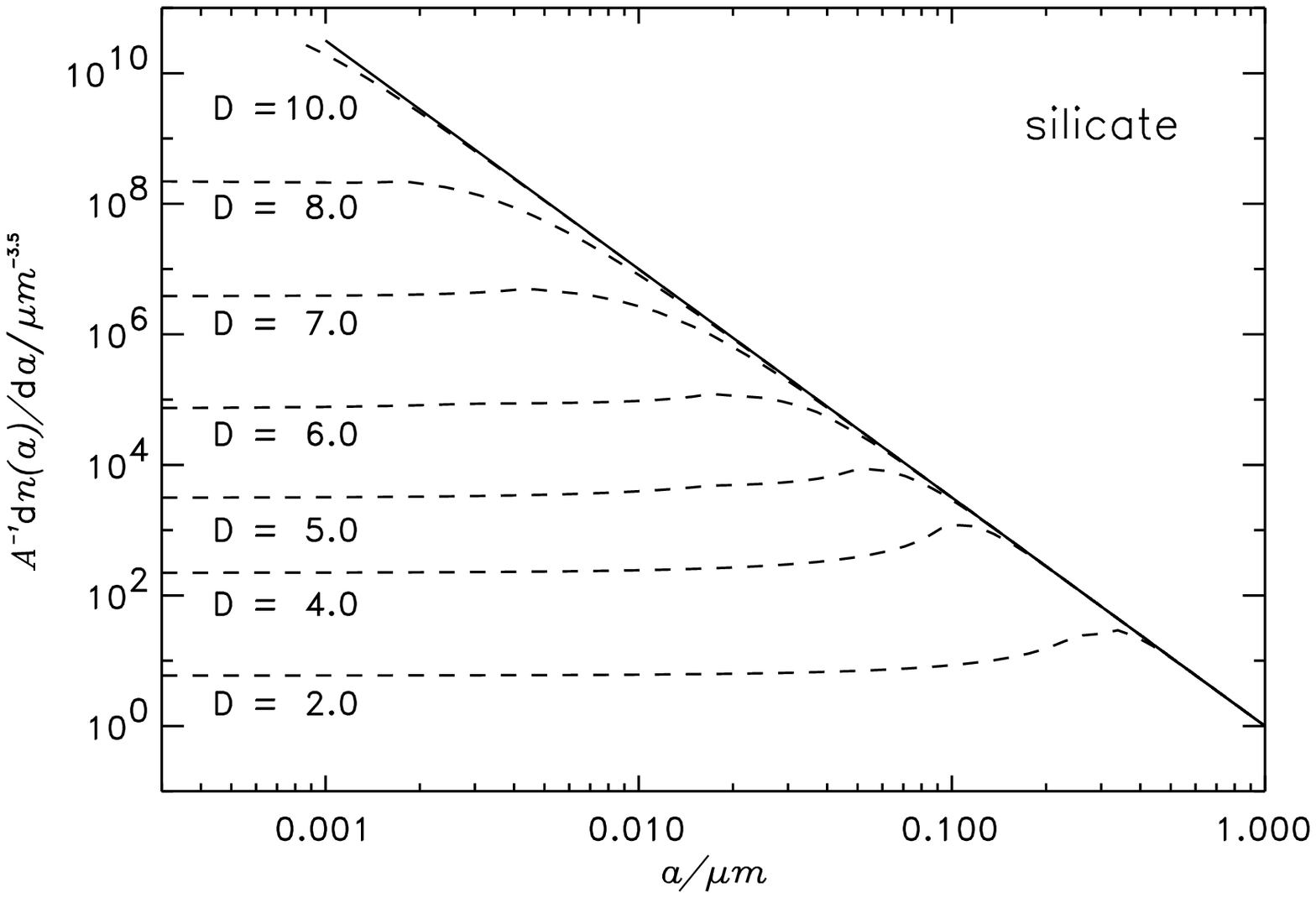}}
  \hfill
  \resizebox{0.49\hsize}{!}{\includegraphics{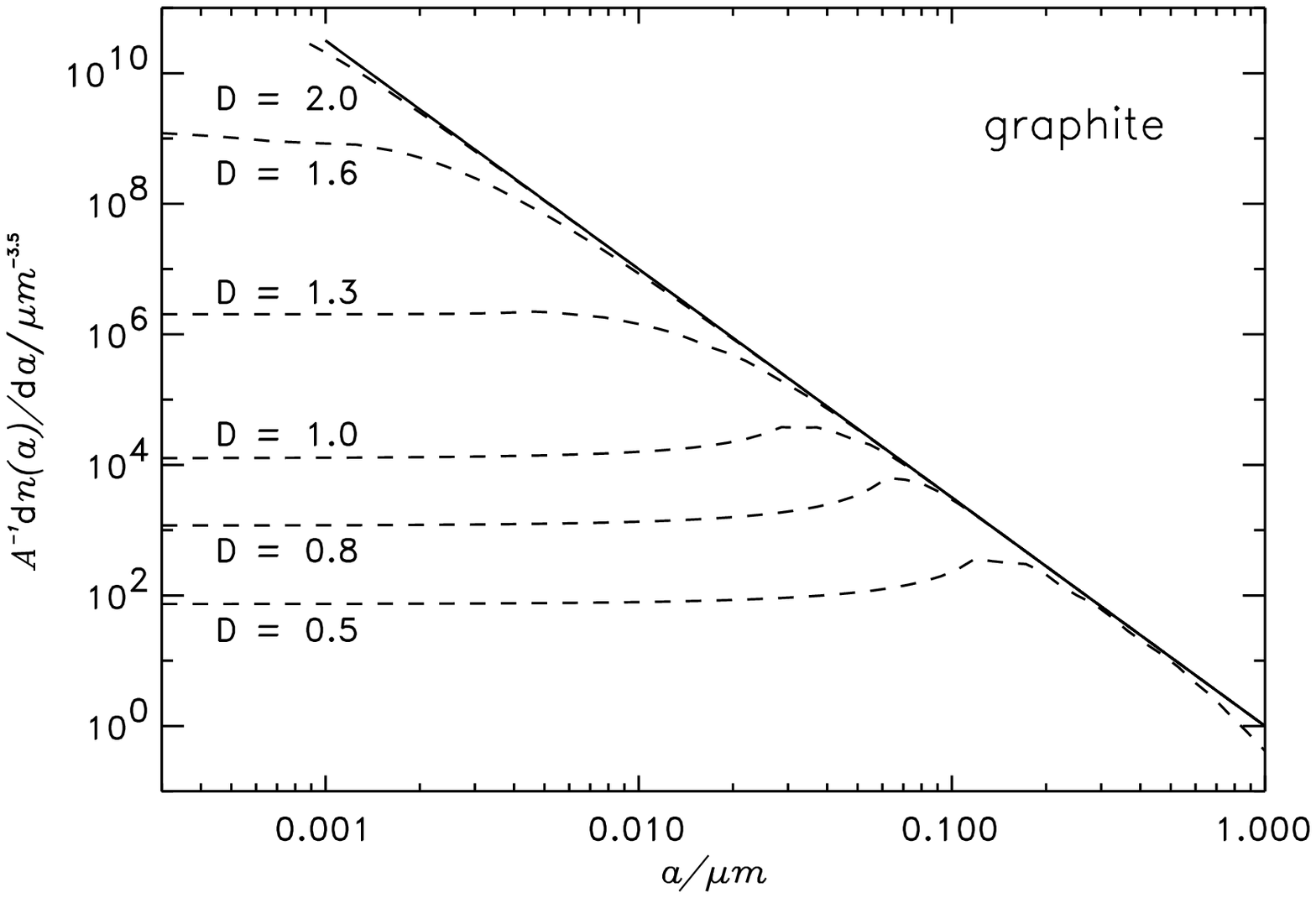}}
  \caption{
   \label{grsizeUV}
    Theoretical grain size distributions (dashed lines) of survived silicate and graphite grains after the
    UV-flash at various distances D (given in units of $10^{15}$~m)
to the supernova. 
    The initial size distribution
    is assumed to have been a power law ${\mathrm d}n(a)=A a^{-3.5}\,{\mathrm d}a$ with constant $A$ 
    (straight solid line).}
\end{figure*}

The total reduction $a_{\rm ev}(a,t)$ of a grain with an initial radius $a$ during the UV-flash
after a time $t$ is found by integrating Eq.~\ref{evap}. If the final radius 
$\tilde a(a,t)=a-a_{\rm ev}(a,t)$  
is smaller than \mbox{3 \AA{}}, the grain is assumed to have evaporated. The final grain size distribution
of non evaporated grains assumed to have an initial size distribution 
${\mathrm d}n(a)=f(a)\,{\rm d}a=A\,a^{-3.5}\,{\mathrm d}a$ with constant A given by:
\begin{eqnarray}
  \tilde{f}(\tilde a)\,{\mathrm d}\tilde a & = & 
  f(a(\tilde a))\,\frac{{\mathrm d}a}{{\mathrm d}\tilde a}\,{\mathrm d}\tilde a \nonumber \\ 
  & = & A\,(\tilde a +a_{\mathrm ev})^{-3.5}
  \left(1-\frac{{\mathrm d}a_{\mathrm{ev}}}{{\mathrm d}a}\right)^{-1}{\mathrm d}\tilde a.
\end{eqnarray}

The derived size distributions of silicate and graphite grains for different distances are shown
in Fig.~\ref{grsizeUV}. 
As the evaporation is a surface effect, the fractional change in radius of grains at the same
temperature is larger for smaller grains.
Therefore the number of small grains decreases 
much more rapidly than the number of bigger grains. In addition, at the same distance to the supernova
smaller grains are generally heated to higher temperatures. Due to the exponential dependence
of Eq.~\ref{evap} the evaporation of smaller grains is faster, which causes the maxima in the 
distribution functions.

\begin{figure}[htbp]
  \resizebox{\hsize}{!}{{\includegraphics{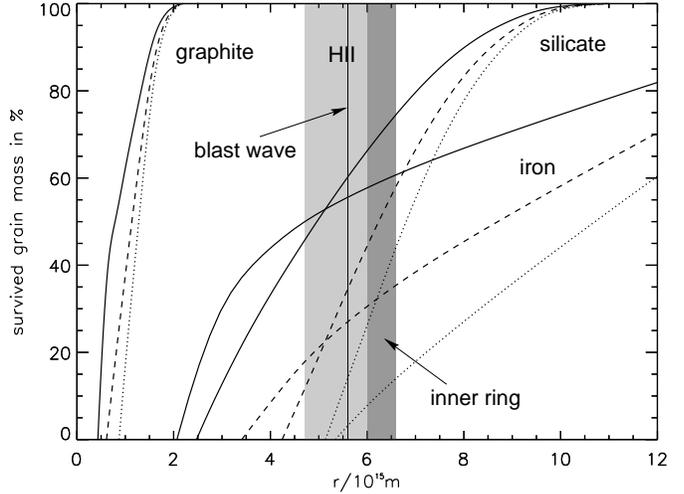}}}
   \caption{ \label{massloss_evap} Survived grain mass of graphite,
   silicate and iron grains after the UV-flash versus distance $r$ from the supernova.  
   It is  assumed, that the
   grain size distribution before the supernova outburst had
   $k=3.5$. $a_{\rm min}$ is chosen to have been $10~${\AA}.  
   The three curves (solid, dashed and dotted
   line) for each dust composition correspond to different choices
   for the initial maximum  grain size $a_{\rm  max}$ ($0.25,$
   $0.1$ and $0.06~\mu{\rm m}$). Also shown is the approximate range of positions
   covered by the thick inner ring (dark grey) and the HII-region (light grey) 
   before being caught by the blast wave.
   The inner radius of the HII-region was taken to be $4.72\times 10^{15}~{\rm m}$.
   This was calculated from the observed position of the blast wave at epoch 3200 days 
   (Gaensler et al., \cite{Gaensler97}) extrapolated to the epoch at which the
   shock first reached the HII-region (1200 days after outburst; 
   Staveley-Smith et al., \cite{Staveley-Smith92}). The vertical straight line at
   a radius of $5.43\times 10^{15}~{\rm m}$
   indicates the position of the blast wave at the epoch of the ISOCAM observations
   (4000 days after outburst, also extrapolated from the observed position at 3200 days).
   The extrapolations were made assuming a velocity of 2900~km/s for the blastwave
   corresponding to model~II.
   If we take the velocity of $4100~{\rm km/s}$ (model I)
   the inner boundary is closer to the supernova
   and the position of the blast wave further out.
   }
\end{figure}

How much silicate, iron and graphite dust might have survived the UV-flash in a certain distance
to the supernova is shown in Fig.~\ref{massloss_evap} for three different maximum grain sizes 
(0.25, 0.1 and $0.06~\mu{\rm m}$). 
It can be seen that silicate grains evaporate out to larger distances from
the supernova than graphite grains. This is partly due to the higher bounding energy of graphite grains
but mainly caused by the much higher temperatures the silicate grains attain in comparison with graphite grains. 
This is the opposite of the situation for collisionally heated dust in the CSM.

The grains most probably responsible for the measured infrared fluxes originated from
between the original inner boundary of the HII-region and the position
of the blast wave 4000 days after outburst. 
The mass loss of iron and silicate
grains in this region is comparable and significant even for the largest considered
grain size of $0.25~\mu{\rm m}$.
This changes at larger distances from the supernova,
where evaporation of predominantly small iron grains becomes much stronger. 
For graphite grains evaporation is unimportant for the whole HII-region. 

The radially integrated evaporated dust masses of silicate, graphite and iron grains corresponding
to the shown curves is given in Table~\ref{masslosstable}. 
In the integration it is assumed that the shock surface area is proportional to the square 
of the distance $r$ of the blast wave to the position of the supernova.
The differences in the derived values for the two models
are due to the different shock velocities, that give slightly different positions
of the inner boundary and the outer shock.

\subsection{Sputtering in the shocked gas}

\label{sputteringsection}
Behind the blast wave grains 
undergo sputtering. This is thought to be one of the most important destruction processes in
fast moving shocks (see e.g. Dwek et al., \cite{Dwek96}). Sputtering time scales
appropriate for a plasma with the abundances of the shocked CSM of SN~1987A 
are given in appendix~\ref{sputteringapp}.

The final grain size distribution after a time $\Delta t$ of an initial grain
size distribution $f(a)\,{\rm d}a$ due to sputtering in a hot plasma is given by:
\begin{equation}
  \tilde f(\tilde a, \Delta t)\,{\mathrm d}\tilde a = 
  f(a(\tilde a,\Delta t))\,\frac{{\mathrm d}a}{{\mathrm d}\tilde a}\,{\mathrm d}\tilde a=
  f(\tilde a+\Delta a(\Delta t))\,{\mathrm d}\tilde a.
\end{equation}
Because of the evaporation of grains during the UV-flash, the grain size distribution
$f(\tilde a)\,{\rm d}a$
entering the sputtering zone is dependent on the distance to the supernova.
The average grain size distribution at the time $t$ after the shock reached the HII-region at time $t_0$
is therefore proportional to:
\begin{equation}
   \left<\tilde f(\tilde a,t-t_0)\right>\propto \int_{t_0}^{t}{\mathrm d}t'\,v_{\rm S}\,\tilde 
   f(\tilde a,t-t',r(t'))\,r(t')^2,
\end{equation}
where $t_0=1200$ days and the surface of the outer shock front is again assumed to 
increase with the square of the distance $r(t)$ of the blast wave to the supernova position.

The derived mass losses for the considered dust compositions and the three different 
maximum grain sizes are tabulated in Table~\ref{masslosstable}. For comparison, we derived
the sputtered dust mass with and without previous evaporation. 
\begin{table}[htbp]
\begin{minipage}[t]{\hsize}
\setlength{\tabcolsep}{1.mm}
\renewcommand{\footnoterule}{\rule{0mm}{0.mm}}
  \caption{
        \label{masslosstable}
        Mass loss through evaporation and sputtering}
  \begin{tabular}{ll|c c c |c c c}
    \hline
    \multicolumn{2}{c|}{$n_{\mathrm{H}}$\footnote{Density of the shocked gas downstream of the blast wave.}} & 
          \multicolumn{3}{c|}{$300\,\mathrm{cm^{-3}}$} & 
      \multicolumn{3}{c}{$600\,\mathrm{cm^{-3}}$} \\
      \hline
      \multicolumn{2}{c|}{$a_{\mathrm{max}}[\mu{\mathrm{m}}]$} & 0.25 & 0.10 & 0.06 & 0.25 & 0.10 & 0.06 \\
      \hline
    $\Delta M_{\mathrm{sputt.}}$\footnote{Sputtering without previous evaporation.} & 
          sil.  & 27.7\% & 41.1\% & 50.3\% & 44.0\% & 61.7\% & 71.9\% \\
    $\Delta M_{\mathrm{evap.}}$ & sil.   & 49.4\% & 80.4\% & 98.3\% & 48.7\% & 79.2\% & 98.9\% \\
    \hline
    $\Delta M_{\mathrm{sputt.}}$ & iron  & 27.2\% & 40.4\% & 49.5\% & 40.9\% & 58.1\% & 68.4\% \\
    $\Delta M_{\mathrm{evap.}}$ & iron   & 48.0\% & 78.6\% & 99.8\% & 47.7\% & 78.1\% & 100.\% \\
    \hline
    $\Delta M_{\mathrm{total}}$\footnote{Mass loss after evaporation and following sputtering.} & 
             sil.   & 53.1\% & 82.4\% & 98.4\% & 58.4\% & 85.1\% & 99.1\% \\
    $\Delta M_{\mathrm{total}}$ & iron   & 51.8\% & 81.0\% & 99.8\% & 56.4\% & 83.6\% & 100\% \\
    $\Delta M_{\mathrm{total}}$ & gra.\footnote{For graphite only the total mass loss is given
        because of an insignificant mass loss due to evaporation.}   
          & 11.3\% & 17.3\% & 22.0\% & 19.1\% & 29.0\% & 36.3\% \\
      \hline
   \end{tabular}
 \end{minipage}
\end{table}
The mass loss due to sputtering becomes progressively more important for the bigger grains.
Whereas sputtering reduces the radius independent of the grain size,
evaporation is the dominant destruction process for small grains.
Graphite grains which are relatively stable
against sputtering (appendix \ref{sputteringapp}) and did not evaporate during the UV-flash 
(Sect. \ref{evaporationsection}) suffer only moderate depletions compared to silicate
and iron grains. In model~I the initial mass of graphite grains would be less than a factor of
1.3 higher than that infered from the ISOCAM observations for all considered maximum grain sizes.




\section{Discussion}

\subsection{Dust in the pre supernvova CSM}

The modelling of the ISOCAM measurements (Sect.~\ref{resultsection}), combined with
the calculations of grain destruction due to evaporation and sputtering 
(Sect.~\ref{destr_section}) allow us to infer grain abundances
in the pre supernova CSM as well as to make crude estimates of maximum grain sizes
and grain composition. 

\begin{table}
\setlength{\tabcolsep}{1.8mm}
\renewcommand{\footnoterule}{\rule{0mm}{0.mm}}
\begin{minipage}[]{\hsize}
  \caption{\label{presndusttable} Solutions for dust in the pre-supernova CSM}
  \begin{tabular}{cc|ccc|ccc}
     \hline
    model & comp. & \multicolumn{3}{c|}{mixture\footnote{Relative mass of 
        silicate to iron or silicate to graphite.}
      } & \multicolumn{3}{c}{$10^{-4}\,M_{\rm d}/M_{\rm gas}$} \\
     & & 0.25\footnote{Maximum grain size in microns.} & 0.10 & 0.06 & 0.25 & 0.10 & 0.06 \\
     \hline
    I     & gra.     & \multicolumn{3}{c|}{--}  & 1.1 & 1.2 & 1.3 \\
    I     & si.+iron & 2.15 & 3.13 & 0.44 & 2.3 &  5.4 & 152 \\
    II    & si.+iron & 2.45 & -- &  -- & 2.8 & -- & -- \\
    I     & si.+gra. & 0.47 & 2.02 & 25.4 & 1.4 & 2.4 & 20. \\
    II    & si.+gra. & 1.42 & 4.01 & 62.3 & 1.8 & 3.6 & 48. \\
   \hline
  \end{tabular}
\end{minipage}
\end{table}

This information was derived for the best solutions ($\chi^2_{\nu}<3$, Table~\ref{results-dustmodel}) using
the mass loss estimates given in Table \ref{masslosstable} 
and is tabulated in Table \ref{presndusttable}. 
As expected the
main parameter determining the composition and abundance of grains in the pre-supernova
CSM is the maximum grain size. It is interesting to compare these quantities with 
observed and derived properties of circumstellar dust in stellar outflows.


We are not aware of direct measurements 
in the IR and submm regimes of dust abundances in the winds of LMC stars.
We will simply estimate it by assuming that the dust abundance in the winds
is proportional to the metallicity of the ISM. This is supported by
Woods et al. 
  (\cite{Wood92}) who found
  that the outflow velocity of oxygen rich stars in the LMC is significantly lower than for oxygen rich stars
  in our galaxy and suggested that the lower velocity is due to the lower dust-to-gas ratio
  caused by the lower metallicity of the LMC. The same has also been
  proposed for high luminosity stars in the galactic anticentre which
  show only a modest outflow velocity (Habing et al., \cite{Habing94}). 
These works were confirmed by van Loon
(\cite{vanLoon00}), who found by comparing obscured Asymptotic Giant Branch
stars of our galaxy, the Large and the Small Magellanic Clouds that the
inferred dust-to-gas ratio of both carbon and oxygen rich stars is approximately proportional 
to the initial metallicity. 

The winds of evolved carbon rich stars in our galaxy have generally been found to have dust-to-gas ratios in
the range $\sim 0.1$ to $\sim 1\%$ (Jura, \cite{Jura86}; Martin \& Rogers, \cite{Martin87}; Griffin, \cite{Griffin90};
Knapp et al., \cite{Knapp93}; Bagnulo et al., \cite{Bagnulo95}; and Olofsson et al., \cite{Olofsson93};
as quoted by Hiriart \& Kwan \cite{Hiriart00}). On the other hand Hiriart \& Kwan (\cite{Hiriart00}) estimated
from a subsample of Olofsson et al. a maximum dust-to-gas ratio of only $\sim 0.1\%$. 
For the galactic oxygen rich star OH~231.8+4.2, Knapp et al. (\cite{Knapp93}) obtained a 
dust abundance of $\sim\,$0.7\%.

Scaling by the factor of 2 between the metallicity of the LMC and the metallicity of the
solar vicinity (Russell \& Dopita, \cite{Russell92}) 
we would expect on this basis dust to gas ratios for LMC carbon stars
in the range $\sim\,0.05\,-\,0.5\,\%$ and for LMC Oxygen stars of order $\sim\,$0.3\%.
Although clearly very uncertain, these numbers are nevertheless only consistent with a subset of the
solutions of the pre-supernova gas-to-dust ratios in the CSM of SN~1987A given in Table~\ref{presndusttable}. 
In particular, all the pure graphite solutions appear underabundant in dust by an order of magnitude
compared with expectations for the winds of carbon stars.
This confirms the expectation that the dust should in fact be silicate rich
on the basis of the gas phase abundances in the inner ring 
(Lundqvist \& Fransson, \cite{Lundqvist96}; Sonneborn et al., \cite{Sonneborn97}).

The solutions for silicate-iron and silicate-graphite mixtures in Table~\ref{presndusttable} are
consistent with the expected dust to gas ratios of $\sim\,0.3\%$ for oxygen-rich stars provided
the maximum grain size in the CSM of SN~1987A was smaller than $0.1~\mu{\rm m}$.
This is in accordance with findings, both theoretical and observational, that
cool stars eject mainly small grains, irrespective of composition.
An upper limit of $0.14~\mu{\rm m}$ on the maximum sizes of grains  
around oxygen rich mass-losing stars was found by (Jura, \cite{Jura96}). 
A maximum grain smaller than $\sim 0.1~\mu{\rm m}$ is also consistent with
the sizes found for the dust ejected from the carbon rich star
IRC+10216 (Martin \& Rogers, \cite{Martin87}; Griffin, \cite{Griffin90}; 
Jura, \cite{Jura94}; Bagnulo et al., \cite{Bagnulo95})
or the typical sizes around evolved carbon stars derived by Hiriart \& Kwan (\cite{Hiriart00}).




On the other hand, this small maximum grain size contrasts with 
results for oxygen rich stars derived from their ultraviolet extinction. 
Rogers, Marting \& Crabtree (\cite{Rogers83}) suggested that the grains
around $\mu$~Cep should be in the range 0.1~$\mu{\rm m}$  to
0.5~$\mu{\rm m}$. Based on extinction measurements made for $\alpha$\,Sco, 
Seab \& Snow (\cite{Seab89}) concluded that the grains in the
circumstellar environments of cool oxygen rich giants should be larger than $\sim
0.08~\mu{\rm m}$ and possibly enrich the ISM with grains as large as $\sim 1~\mu\mathrm{m}$. 
If this were also the case for the RSG wind of the progenitor of SN~1987A, 
the low dust abundance found by ISOCAM could neither be explained by evaporation nor by 
sputtering. It might then have had to have been intrinsically low.
Alternatively, the grain abundance could have
been reduced by radiation pressure (Turner \& Pearce, \cite{Turner92})
after the progenitor evolved to
its final BSG phase 20\,000 years ago (Crotts \& Heathcote, \cite{Crotts91}).
However for this, the magnetic coupling of the grains to the gas
would have had to have been weak. Another scenario for reducing the grain
abundance might be a mixing of the material of the BSG wind, with a 
negligible dust abundance, with that of the RSG wind at a dynamically unstable
interface between the winds (Garcia et al., \cite{Garcia96a}, \cite{Garcia96b}).

\subsection{Iron abundance in the shocked CSM}

On the basis of a spherically symmetrical hydrodynamical simulation 
of the interaction of the blast wave with the HII region, 
Borkowski et al. (\cite{Borkowski97}) derived an upper limit on
the gas phase iron abundance in the HII region
of only 0.1 of the solar iron abundance from the X-ray spectrum 
measured with the ROSAT satellite (Hasinger et al., \cite{Hasinger96}).
Comparing Chandra data with a plane parallel shock model Park et al. (\cite{Park02}) found
a value of 0.07 of solar for the iron abundance in the X-ray emitting region
between epochs 1999 and 2001.
These gas phase abundances for iron are lower by a factor of at least $\sim 3$
compared to the iron abundance in the LMC from Russell \& Dopita (\cite{Russell92}), 
which is $36\,\%$ of the solar abundance (determined from the photosphere; 
Anders\& Grevesse 1989). This prompted 
Borkowski et al. to suppose that most of the iron was
condensed into grains. 
However, this is not supported by the ISOCAM measurements.
An upper limit for the mass of iron condensed in grains can be 
taken from the calculations for pure iron grains in the shocked
CSM tabulated in Table~\ref{results-dustmodel}. Taking for iron grain masses
of $1.5\times 10^{-6}\,\mathrm{M_{\sun}}$ (model I) or 
$1.1\times 10^{-6}\,\mathrm{M_{\sun}}$ (model II) and the
iron abundance in the LMC to be $n_{\rm Fe}/n_{\rm H}=1.7\times 
10^{-5}$ (Russell \& Dopita, \cite{Russell92}), the  fraction of iron 
in solid form is at maximum $\sim 32\%$ (model I) or
$\sim 45\%$ (model II). For the better fitting silicate/iron
mixtures the upper limits will be still lower.
On the basis of this evidence, the HII-region would underabundant in iron, 
whether in gaseous or condensed form. Further X-ray and
infrared observations would be valuable to investigate this
problem.


\subsection{Outlook}
In the short term, further observations to follow the MIR light curve will 
provide information on the dependence on radial position of the dust-to-gas ratios in
the shocked HII region for comparison with model predictions. 
On the one hand there will be a tendency for the  overall volume-averaged dust-to-gas ratios 
to be lowered due to sputtering at later epochs, if the upstream dust abundance is constant.
On the other hand, the survived dust abundance after the UV-flash will increase with radius,
especially for silicate and iron grains. Another
potential reason for increasing dust abundances with radius could be a flushing out of
grains from the inner regions of the CSM through radiation pressure, after the RSG
turned into a BSG. This might also offer an explanation for the puzzle of the low iron abundance 
in both grains and gas inferred from the ISOCAM and ROSAT and CHANDRA data, as discussed above. 
As already stated this may require a weak magnetic coupling of the grains to the gas.


Observations of the brightening so-called ``hot spots'' seen in the optical (Lawrence et al., \cite{Lawrence00}) 
show that the blast wave is already interacting at certain places with dense material of the thick inner ring.
The origin of the thick inner ring, is unknown. It is also not clear whether the MIR echo seen after 580 days
(Roche et al., \cite{Roche93}) can be attributed to the ring. If the ring is composed of material from the
red supergiant phase of the progenitor star (see e.g. Fransson et al., \cite{Fransson89}), then one might 
anticipate a rapid increase in MIR luminosity with time, which would soon dominate 
the continuum emission from the HII-region. The higher gas densities in the ring might lead to
an accompanying increase in grain temperature, which would allow 
thermal dust emission from the thick inner ring to be distinguished from an increasing
contribution from the HII region. If, on the other hand, the thick inner ring is composed of
material from the companion star in the putative binary system (e.g. Podsiadlowski, \cite{Podsiadlowski92})
then one might speculate that the dust abundance and composition of the ring might deviate markedly
from that of the HII region discussed in this paper.

\section{Summary}

The IR emission was analysed in terms of thermal emission from dust, collisionally heated by the
shocked gas behind the blast wave that is expanding into the HII-region interior to the thick 
inner ring. We used a realistic grain model with a grain size distribution 
$\mathrm{d}n\propto a^{-k}\mathrm{d}a$ including
stochastically heating of small grains. For the shocked
gas we considered two different models corresponding to the downstream densities $n_{\rm H}=300~{\rm cm^{-3}}$
(model I) and $n_{\rm H}=600~{\rm cm^{-3}}$ (model II). The conclusions for the shocked CSM are as
follows:
\begin{enumerate}
  \item The luminosity of the dust emission is found to range from  $2.91$ to $3.35\times 10^{28}$~W 
    (according to assumed grain composition and plasma density).
  \item 
    The dust is most likely composed either of a silicate-iron or a silicate-graphite mixture
    or of pure graphite. 
  \item Considering only cases with a maximum grain size not larger
than $0.25~\mu{\rm m}$ 
    the dust masses are, independent of the assumed grain composition and size distribution, 
    $(0.9~\mathrm{to}~ 1.1)\times 10^{-6}~\mathrm{M_{\sun}}$ (model I) and 
    $(0.45~\mathrm{to}~0.67)\times 10^{-6}$ (model II).
    This corresponds for a gas mass of $1.0\times 10^{-2}/n_{\rm H}[300~{\rm cm^{-3}}]~\mathrm{M_{\sun}}$ 
	to dust-to-gas ratios in the range $(0.9~\mathrm{to}~1.3)\times 10^{-4}$. 
    For larger maximum grain sizes a larger range of the dust-to-gas
ratio of up to
    $2.2\times 10 ^{-4}$ are admitted by the data.
  \item For the LMC abundances of Russell \& Dopita (\cite{Russell92}) the maximum
    fraction of iron condensed into grains in the shocked CSM is $32\%$ (model~I) or $45\%$ (model~II).
\end{enumerate}
We have calculated the grain destruction due to the UV-flash of the supernova outburst and
subsequent sputtering downstream of the blast wave. From this we have deduced dust properties
in the CSM prior to the supernova. Our conclusions are:
\begin{enumerate}
  \item Pure graphite solutions for the pre-supernova CSM are underabundant
    in dust by an order of magnitude compared with estimates for the abundance
    of carbon grains in the winds of LMC-carbon stars.
  \item Solutions for silicate-iron and silicate-graphite mixtures in the
    pre-supernova CSM are consistent with the expected dust-to-gas ratios of
    $\sim 0.3\%$ in the winds of oxygen rich LMC stars provided the maximum
    grain size was less than $\sim 0.1~\mu{\rm m}$.
\end{enumerate}

\begin{acknowledgements}
The work was supported by Deutsches Zentrum f\"ur Luft- und Raumfahrt e.V. (DLR) 
through the projects `50 OR 9702' and `50 OR 99140'. 
We have made use of the {\em ROSAT} Data Archive of the Max-Planck-Institut 
f\"ur extraterrestrische Physik (MPE) at Garching, Germany.
\end{acknowledgements}

\appendix

\section{X-ray flux until 4000 days after outburst}

\label{x-ray-flux}

Measurement of the X-ray flux from the SN 1987A 
until $\sim 3000$ days after outburst with the X-ray satellite ROSAT
(Hasinger et al., \cite{Hasinger96}) showed a monotonic increase of the luminosity, where
the variation in time could be linear or also steeper with $\propto t^{1.67\pm 0.35}$.
A linear trend would be consistent with a constant external density (Hasinger et al., \cite{Hasinger96}).
Later measurements with CHANDRA $\sim 13$ years after outburst gave fluxes clearly
above a linear trend (Burrows et al., \cite{Burrows00}). 
To see how the X-ray flux evolved until the ISOCAM observations we derived the X-ray fluxes
from a large number of available measurements that where made with the HRI and the PSPC instrument
of ROSAT until $\sim 4000$ days after outburst. For a direct
comparison with the data published by Hasinger et al. we also included
measurements which were made before $\sim 3000$ days after outburst. The
data were automatically calibrated with SASS. In calculating the
counts of X-ray photons we used the same apertures to determine the
source and the background counts used by Hasinger et al.
(\cite{Hasinger96}) and scaled the photon fluxes derived from the
HRI-data with a factor 2.65 to allow comparison with the PSPC-data. To
estimate the uncertainties we assumed poisson statistics for the
source counts and gaussian statistics for the background noise.

\begin{table}[tbph]
\renewcommand{\footnoterule}{\rule{0mm}{0.mm}}
\begin{minipage}[]{\hsize}
        \caption[]{
        Derived rates of X-ray photons towards SN~1987A.
        \label{rosat_flux}
        }
  \begin{tabular}{c|ccrc}
     \hline
     obs. time & day & instr.\footnote{Number in brackets give the number of observations used.} 
        & $t_{\rm int}/{\rm s}$ & N/1000\,s      \\
        \hline
        12.02.91 - 13.02.91 &1448& HRI (1) & 23107 & 0.17 $\pm$ 0.90   \\
        06.10.91 - 07.10.91 &1685& PSPCB & 16398 & 2.72 $\pm$ 0.63 \\
        30.04.92 - 14.05.92 &1898& PSPCB &9340& 2.05 $\pm$ 0.76 \\
        04.12.92 - 06.12.92 &2110& PSPCB &2552& 5.12 $\pm$ 1.80 \\
        07.04.93 - 10.04.93 &2235& PSPCB &11259& 2.63 $\pm$ 0.73 \\
        20.06.93 - 05.07.93 &2315& PSPCB &10391& 3.13 $\pm$ 0.76 \\
        28.09.93 - 30.09.93 &2409& PSPCB &9131& 3.79 $\pm$ 0.85 \\
        20.06.94 - 20.09.94 &2718& HRI (4)&12223& 5.01 $\pm$ 1.62 \\
        03.10.94 - 02.01.95 &2823& HRI (2)&18756& 3.54 $\pm$ 0.91 \\
        01.04.95 - 11.07.95 &3008& HRI (3)&28833& 4.61 $\pm$ 0.97 \\
        10.10.95 - 10.01.96 &3196& HRI (2)&26034& 7.29 $\pm$ 1.15 \\
        16.04.96 - 31.07.96 &3392& HRI (2)&46907& 7.54 $\pm$ 0.89 \\
        22.10.96 - 12.01.97 &3568& HRI (2)&45192& 7.22 $\pm$ 0.87 \\
        21.02.97 - 02.03.97 &3654& PSPCB &15440& 8.31 $\pm$ 0.86 \\
        04.03.97 - 03.06.97 &3706& HRI (2)&52601& 8.31 $\pm$ 0.86 \\
        16.12.97 - 17.12.97 &3948& HRI (1)&21738& 12.6 $\pm$ 1.5 \\
        19.02.98 - 22.02.98 &4014& PSPCB &19382& 9.19 $\pm$ 0.78 \\
  \hline
  \end{tabular}
\end{minipage}\end{table}

\begin{figure}
  \resizebox{0.99\hsize}{!}{\includegraphics{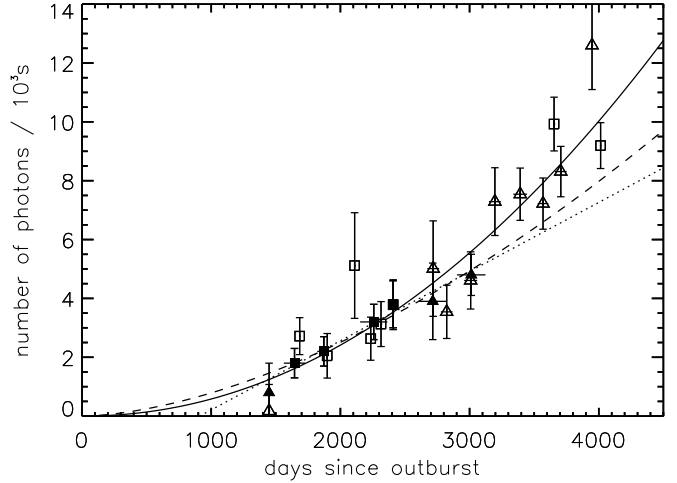}}
  \caption[]{
    \label{xrayflux}
    Time evolution of the X-ray emission of SN 1987A, measured with the {\em ROSAT} satellite. 
    The photon fluxes derived from observations with the HRI and the PSPC instrument 
    are shown as triangles and squares. Published data  (filled symbols) 
    until 3000 days after outburst
    (Hasinger et al., \cite{Hasinger96}) are shown for comparison. 
    These data can be described through a
    linear increase (dotted line) since 900 days and a potential (broken line) 
    with $\propto t^{1.67}$ (Hasinger et al., \cite{Hasinger96}).
    The photon fluxes derived here (open symbols) give an increase 
    with $\propto t^{2.06\pm0.20}$ (solid line). 
    }
\end{figure}

The derived final counting rates of the observations are summarised in table \ref{rosat_flux} and
shown in Fig.~\ref{xrayflux} as open symbols. Until 3000 days after
outburst they are consistent with the counting rates published by
Hasinger et al. which are shown as black solid symbols.  At later
times the fluxes lay slightly above the linear trend. Fitting a potential
to the counting rates derived here gives with a $\chi^2_{\rm min}=1.29$ a
monotonic increase of: 
\begin{equation}
  N(t) \propto t^{2.06\pm 0.20}.
\end{equation}
This may be because the shock is reaching denser material in time. 
On the other hand, numerical calculations
for a HII-region with homogeneous density do not give 
a linear trend of the X-ray luminosity for the first several years 
(Borkowski et al., \cite{Borkowski97}).

To derive the volume $V$ of the X-ray emitting gas at the time of the ISOCAM observations $\sim 4000$ days 
after outburst we assumed that the X-ray luminosity was proportional to the 
emission measure $EM=n_{\rm e}n_{\rm i}V$, where $n_{\rm e}$ and $n_{\rm i}$ are the number density of the electrons
and the ions. Taking the emission measure  after $\sim 2500$ days to be 
$(1.4\pm 0.4)\times 10^{57}~{\rm cm^{-3}}$ (Hasinger et al., \cite{Hasinger96}; Borkowski et al., \cite{Borkowski97}) 
and the abundances as given in Sect.~\ref{parametergas},
this volume was approximately $2.2\times 10^{52}~{\rm cm^3}$.

\section{Sputtering time scales for the shocked CSM}

\label{sputteringapp}
Here we derive sputtering rates specific to the physical conditions and abundances
encountered in the shocked HII region around SN~1987A.
At the high temperatures involved, 
sputtering is independent of grain charge and nearly independent of the plasma temperature.
Sputtering reduces the radius of all grains of a given composition 
in a time $\Delta t$ by the same amount $\Delta a(\Delta t)$.
To derive the life times of the grains due to sputtering we used the formula (Eq.~27) given by Draine \& Salpeter 
(\cite{Draine79a}) for non rotating grains moving with a relative velocity $v_{\rm gr}$ 
with respect to the gas with a temperature $T_{\rm i}$. Sputtering yields were
adopted from Tielens et al. (\cite{Tielens94}), and we considered sputtering due to
H, He, C, N and O, adopting plasma abundances as given in Sect. \ref{parametergas}.
The initial relative velocities of grains overtaken by the blast wave ($v_{\rm gr}=\frac{3}{4}v_{\rm S}$) are so high
that at first the sputtering is almost non thermal. Due to the drag forces the sputtering becomes 
thermal after some time, which increases the sputtering yield slightly (10\%-30\%). Here we simply take the
average of the sputtering yield of the two limits $v_{\rm gr}=0$ and $v_{\rm gr}=3v_{\rm s}/4$. The
resulting life times of the considered grains are:
\begin{equation}
  \tau = \frac{a[0.01\mu{\mathrm{m}}]}{n_{\rm H}[{\rm cm^{-3}}]}\times\cases{\begin{array}{rl}
        11.5 \times 10^3 & \quad \mbox{graphite}\\
        2.3  \times 10^3 & \quad \mbox{silicate}\\
        2.7  \times 10^3 & \quad \mbox{iron}
        \end{array}} \quad \mathrm{years}
\end{equation}
in model I and 
\begin{equation}
  \tau = \frac{a[0.01\mu{\mathrm{m}}]}{n_{\mathrm{H}}[\mathrm{cm^{-3}}]}\times\cases{\begin{array}{rl}
        12.8 \times 10^3 & \quad \mbox{graphite}\\
        2.9  \times 10^3 & \quad \mbox{silicate}\\
        3.0  \times 10^3 & \quad \mbox{iron}
        \end{array}} \quad \mathrm{years}
\end{equation}
in model II.


\end{document}